\documentclass[aps,prb,twocolumn,superscriptaddress,showpacs]{revtex4}
\usepackage{amsfonts}
\usepackage{amssymb}
\usepackage{amsmath}
\usepackage{graphicx}

\usepackage{dcolumn}   
\usepackage{bm}        
\usepackage{flafter}

\newcommand{\Tc}{\ensuremath{T_c}}

\begin{document}


\title{Electron-hole asymmetry in Co- and Mn-doped SrFe$_2$As$_2$}

\date{\today}
\author{J. S. Kim}
\email[E-mail~]{js.kim@postech.ac.kr}
\affiliation{Department of Physics, Pohang University of Science and Technology, Pohang 790-784, Korea}
\author{Seunghyun Khim}
\affiliation{FPRD, Department of Physics and Astronomy, Seoul National University,
Seoul 151-742, Korea}
\author{H. J. Kim}
\affiliation{Department of Physics, Pohang University of Science and Technology, Pohang 790-784, Korea}
\author{M. J. Eom}
\affiliation{Department of Physics, Pohang University of Science and Technology, Pohang 790-784, Korea}
\author{J. Law}
\affiliation{Max-Planck-Institut f\"{u}r Festk\"{o}rperforschung,
Heisenbergstra$\rm\beta$e 1, D-70569 Stuttgart, Germany}
\affiliation{Department of Physics, Loughborough University, Loughborough, LE11 3TU, United Kingdom}
\author{R. K. Kremer}
\affiliation{Max-Planck-Institut f\"{u}r Festk\"{o}rperforschung,
Heisenbergstra$\rm\beta$e 1, D-70569 Stuttgart, Germany}
\author{Ji Hoon Shim}
\affiliation{Department of Chemistry, Pohang University of Science and Technology, Pohang 790-784, Korea}
\author{Kee Hoon Kim}
\email[E-mail~]{optopia@snu.ac.kr}
\affiliation{FPRD, Department of Physics and Astronomy, Seoul National University,
Seoul 151-742, Korea}
\email[Corresponding author:~E-mail~]{optopia@snu.ac.kr}

\begin{abstract}
Phase diagram of electron and hole-doped SrFe$_2$As$_2$ single crystals is investigated using Co and Mn substitution at the Fe-sites. We found that the spin-density-wave state is suppressed by both dopants, but the superconducting phase appears only for Co (electron)-doping, not for Mn (hole)-doping. Absence of the superconductivity by Mn-doping is in sharp contrast to the hole-doped system with K-substitution at the Sr sites. Distinct structural change, in particular the increase of the Fe-As distance by Mn-doping is important to have a magnetic and semiconducting ground state as confirmed by first principles calculations. The absence of electron-hole symmetry in the Fe-site-doped SrFe$_2$As$_2$ suggests that the occurrence of high-$\Tc$ superconductivity is sensitive to the structural modification rather than the charge doping.
\end{abstract}
\smallskip


\maketitle

Whether or not an electron-hole symmetry holds for high-$\Tc$ cuprates has been an important issue for understanding the origin of high-$\Tc$ superconductivity. Although the details of the doping dependence are different, both hole and electron doping into the Mott-insulating parent compounds destroy the antiferromagnetic (AFM) ground state and eventually lead to high-$\Tc$ superconductivity. For the newly-discovered high-$\Tc$ Fe-pnictides\cite{LaOFeAs:hosono:syn}, such an electron-hole symmetry seems to be valid as demonstrated by several experiments so far.\cite{FeAs:eh_sym} In spite of the distinct \emph{itinerant} AFM ground state\cite{SDW} of the parent compounds, the phase diagram looks very similar with that of high-$\Tc$ cuprates; superconducting phase boundary with a dome shape is formed at the region where the AFM phase is completely suppressed by both hole and electron doping. Understanding the common phase diagram of cuprates and Fe-pnictides, therefore, is an essential step towards understanding their high-$\Tc$ superconductivity in the vicinity of the AFM phase.

Doping dependence of the electronic structure of Fe-pnictides supports the electron-hole symmetry. Generally, the electronic structure of Fe-pnictides near the Fermi level($E_F$) consists of several bands mainly from Fe 3$d$-orbitals hybridized with As $p$-orbitals.\cite{LaOFeAs:pickett:band,FeAs:singh:band} The undoped compounds shows two different types of the Fermi surface (FS), hole pockets at the $\Gamma$ point and electron pockets at the $M$ point in the Brilouin zone. They have the almost same size, which leads to a spin-density-wave (SDW)-type instability through strong inter-band nesting effects.\cite{FeAs:mazin:theory,FeAs:eremin:theory} Upon hole doping, for example, the hole pockets grow while the electron pockets shrink, which in turn spoils the nesting condition for the SDW phase.\cite{LaOFeAs:pickett:band} Once the SDW phase is suppressed sufficiently, the strong inter-band scattering is believed to provide important pairing channel for superconductivity.\cite{FeAs:mazin:theory,FeAs:Kuroki:theory} For higher doping, however, the electron bands are completely emptied losing the interband pairing channel, and accordingly superconductivity is also suppressed with decrease of $T_c$.\cite{KFe2As2:sato:ARPES} Similar mechanism can be also operating in the electron-doping regime\cite{Ba122_Co:hhwen:hall}, which may lead to the electron-hole symmetry in the phase diagram.

The structure of the Fe-pnictides consists of a common building block, the FeAs layer, and a charge reservoir layer, this being either a $RE$O layer ($RE$ = Rear earths) for the so-called "1111" compounds or an $AE$ layer ($AE$ = Alkaline earths) for the "122" compounds. By modifying the charge reservoir or direct substitution at the Fe-sites, additional charge carriers can be introduced into the FeAs layers. For example, in the 122 compounds, Co,\cite{Ba122_Co:mandrus:syn,Ba122_Co:fisher:syn,Ba122_Co:hhwen:hall,Sr122_Co:geibel:syn,Sr122_Co:jskim:syn,Sr122_CoNiMn:rosner:syn} Ni,\cite{Ba122_Ni:li:syn,Sr122_CoNiMn:rosner:syn,Sr122_Ni:paglione:syn} Pd, Rh, Ir,\cite{Ba122_RhPd:Canfield:syn,Sr122_RhIrPd:hhwen:syn} Pt \cite{Ba122_Pt:paglione:syn} substitution at the Fe-sites causes electron-doping, while K-substitution at the $AE$ sites\cite{Ba122_K:johrendt:syn,Sr122_K:sasmal:syn} or Mn-substitution at the Fe sites\cite{Sr122_CoNiMn:rosner:syn} leads to hole-doping. Among them, Mn-doping shows a quite unique behavior; while all the other types of doing are successful for inducing superconductivity, Mn-doping does not induce superconductivity. These findings suggest that not only the modulation of charge concentration but also other effects have to be taken into account for understanding the effects of doping.

Herein, we present the phase diagram of electron- and hole-doped SrFe$_2$As$_2$ single crystals using direct substitution of Co and Mn at the Fe-sites. Similar to the case of K-doped SrFe$_2$As$_2$ at the Sr-site\cite{Sr122_K:sasmal:syn}, Mn-doping at the Fe-sites suppress the SDW transition. However, Mn-doping does not induce the superconductivity but rather makes the system more magnetic and insulating, revealing the absence of electron-hole symmetry. We found that the structure is deeply related to the electron-hole asymmetry. Structural changes induced by K- and Mn-doping, in particular, the changes of the Fe-As distance turn out to be quite different.
First principles calculation also confirms that Mn-doped system favors the magnetic ground state due to larger Fe-As distance. Our results show that the superconductivity is not induced only by the suppression of the SDW state but also avoiding the strong increase of the Fe-As distance.

Single crystals of Co- and Mn-doped SrFe$_2$As$_2$ were grown using Sn-flux techniques as described in detail elsewhere.\cite{Sr122_Co:jskim:syn} X-ray diffraction (XRD) on single crystals reveals sharp (00$l$) peaks, confirming a successful growth. In order to extract the structural information, we carried out powder XRD experiments using both Cu-K$_{\alpha}$ and Mo-K$_{\alpha}$ radiation with a Debye-Scherrer configuration, on powders made of individual single crystals. No additional diffraction peaks has been detected, indicating good quality of the samples. The Co or Mn concentrations were determined by energy dispersive X-ray spectroscopy. The in-plane resistivity was measured by the standard 4-probe method. Magnetization measurements were done under 7 T magnetic field along the $ab$-plane using a SQUID magnetometer. First principles electronic structure calculations are done by the full-potential linearized augmented plane-wave method implemented in Wien2k code\cite{wien2k}. The local spin density approximation is used for the exchange-correlation interaction. In order to describe the magnetic state, we construct supercell assuming each magnetic ground state with tetragonal structure obtained from experiments. Around 1000 k-points are used for the full Brillouin zone integration.

\begin{figure}
\includegraphics*[width=8.5cm, bb=0 0 592 436]{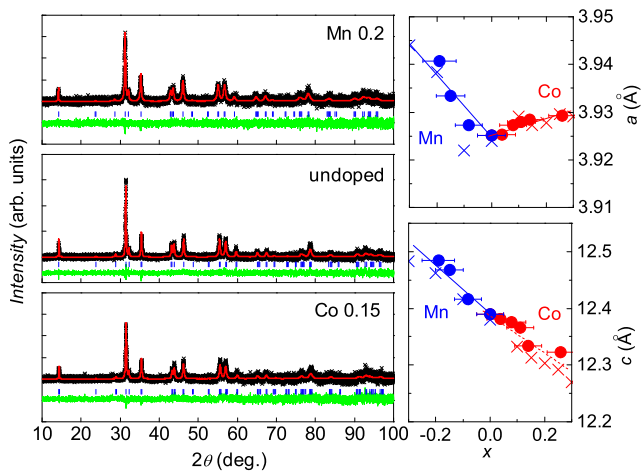}
\caption{\label{fig1}(Color online) Typical X-ray powder diffraction patterns for (a) undoped, (b) Mn-doped ($x$=0.15), and (c) Co-doped ($x$=0.15) SrFe$_2$As$_2$. The red (dark gray) solid line represents the Rietveld refinement (experimental data). The difference between the observed and the calculated data is shown by the solid line below the row of vertical bars marking the angles of the Bragg reflections used to simulate the patterns. Doping dependence of lattice constants (d) $a$ and (e) $c$ for SrFe$_{2-x}$$M_x$As$_2$ (filled circle). For comparison, we also plot those of polycrystalline samples (Ref. [\cite{Sr122_CoNiMn:rosner:syn}]) (cross).}
\end{figure}

Figure \ref{fig1}  shows the typical powder diffraction patterns for undoped and Mn-/Co-doped SrFe$_2$As$_2$. Rietveld profile refinement was performed simultaneously for both XRD patterns taken with Cu-K$_{\alpha}$ and Mo-K$_{\alpha}$ radiation, based on the space group of I4/$mmm$ with the program Fullprof\cite{fullprof}. The converged parameters include the lattice constants, the fractional atomic position coordinates (0,0,$z$) of the As 4$e$ site, and isotropic thermal parameters for all atoms. The $R_p$, $R_{wp}$ and the reduced $\chi^2$ is typically $\sim$ 3, 4 and 1.2, respectively. The $c$-axis lattice parameter shows a linear decrease from Mn- to Co-doping with $\approx$ 0.5 $\AA$ per doping ($x$), consistent with previous results on polycristalline samples.\cite{Sr122_CoNiMn:rosner:syn}  The in-plane lattice constant $a$ slightly increases for both Co- and Mn-doping with a somewhat larger rate for Mn than for Co. Considering the ionic size of Mn$^{2+}$ (0.66 $\rm{\AA}$), Fe$^{2+}$(0.63 $\rm{\AA}$) and Co$^{2+}$ (0.58 $\rm{\AA}$)\cite{TM:shannon:size}, we note that the non-monotonous behavior of $a$ with doping cannot be simply attributed to the different ionic size of the substituted atoms.

\begin{figure}
\includegraphics*[width=8.5cm, bb=0 0 590 400]{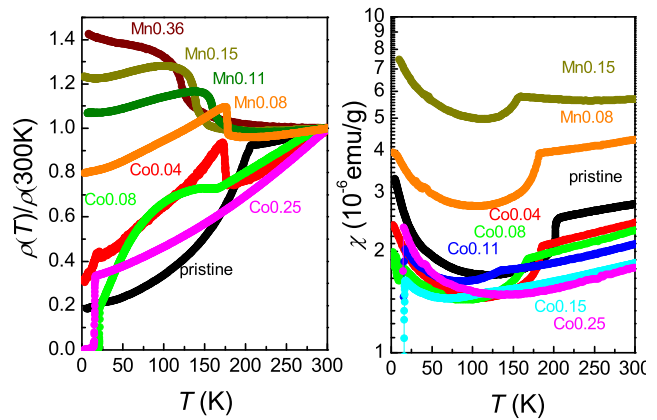}
\caption{\label{fig2}(Color online) Temperature dependence of (a) normalized resistivity $\rho(T)$/$\rho(300K)$ and (b) magnetic susceptibility $\chi(T)$ at $H$ = 7 T along the $ab$-plane for Co- and Mn-doped SrFe$_2$As$_2$.}
\end{figure}

Figure \ref{fig2}  shows the temperature dependence of the normalized in-plane resistivity $\rho(T)$ and the in-plane magnetic susceptibility $\chi(T)$ in SrFe$_{2-x}$$M_x$As$_2$ at $H$ = 7 T along the $ab$-plane. A sudden drop in $\rho(T)$ and $\chi(T)$ in the parent SrFe$_2$As$_2$ at $T_{\rm SDW}$ = 205 K corresponds to the SDW transition. Upon moderate Co-doping, the anomaly in $\rho(T)$ becomes a sharp peak, which then shifts to lower temperatures and smears out with further Co-concentration. The kink in $\chi(T)$ also shifts to lower temperatures with doping, consistent with the behavior of $\rho(T)$. For $x$ $\geq$ 0.04, superconductivity is signaled by an abrupt decrease of $\rho(T)$ near 20 K, and it is fully developed as a zero-resistivity state for $x$ $\geq$ 0.08. Below $x$ = 0.15, we found clear coexistence of the magnetic and superconducting transitions that has not been observed in polycrystalline samples\cite{Sr122_Co:geibel:syn}.

For Mn-doping, on the other hand, the temperature dependences of $\rho(T)$ and $\chi(T)$ are quite different from those of Co-doping. The anomaly in $\rho(T)$ and $\chi(T)$ due to the SDW transition shifts to lower temperatures as in Co-doping, but it becomes more pronounced with doping. Below $T_{\rm SDW}$, $\rho(T)$ shows a much weaker temperature dependence, and upon further doping, $\rho(T)$ becomes almost temperature-independent at low temperatures indicating strong carrier localization. The overall magnitude of $\chi(T)$ increases with Mn-doping while it becomes smaller with Co-doping. Such behavior in $\rho(T)$ and $\chi(T)$ for Mn-doping is also in sharp contrast with the case of K-doping. For K-doping, the anomaly in $\rho(T)$ becomes weaker and $\rho(T)$ shows a more metallic behavior with doping before it eventually drops at the superconducting transition.\cite{Sr122_K:sasmal:syn} These findings clearly demonstrate that Mn-doping suppresses the SDW transition, similar to Co- and K-doping, but drives the system to a distinct magnetic ground state different from the superconducting one.

\begin{figure}
\includegraphics*[width=7.5cm, bb=0 0 600 500]{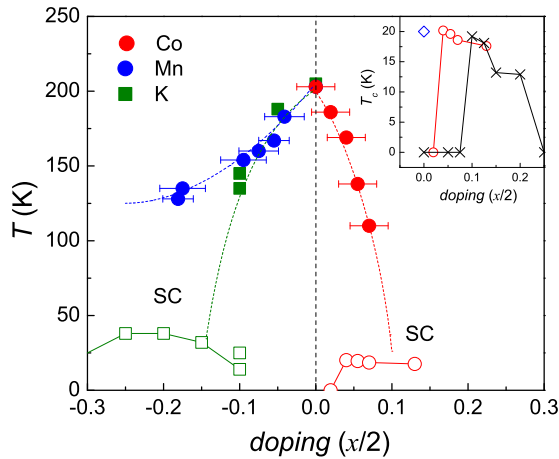}
\caption{\label{fig3}(Color online) Phase diagrams of SrFe$_{2-x}$$M_x$As$_2$ ($M$ = Co and Mn). The solid symbols indicate the AFM transition temperatures, while the open ones represent the superconducting transition temperatures. Note that Mn-doping does not induce the superconductivity while K-dopoing does.(Ref.\cite{Sr122_K:sasmal:syn}) The inset shows the doping dependence of $\Tc$ for Co-doped SrFe$_2$As$_2$ single crystals (open circle). For comparison, together plotted are $\Tc$ of polycrystalline SrFe$_{2-x}$Co$_x$As$_2$ in Ref. \cite{Sr122_Co:geibel:syn} (cross) and strain-induced SrFe$_2$As$_2$ in Ref. \cite{Sr122:paglione:SC} (open diamond).}
\end{figure}

The phase diagram of SrFe$_{2-x}$$M_x$As$_2$ determined from $\rho(T)$ and $\chi(T)$ is presented in Fig. \ref{fig3} . Here, we used the number of extra-conduction electrons per Fe-site with respect to the un-doped system, thus the negative value indicates hole-doping. $T_{\rm SDW}$ was determined as the temperature where $\rho(T)$ begins to increase sharply or $\chi(T)$ shows a kink, which are in good agreement with each other. For comparison, $T_{\rm SDW}$ from K-doped SrFe$_2$As$_2$ is also shown. $T_{\rm SDW}$ decreases monotonically with both type of doping, and the decreasing rate is somewhat larger for electron-doping than hole-doping. The reduction of $T_{\rm SDW}$ can be understood in terms of the weakened FS nesting effect. Since the SDW transition is closely related to the instability due to strong FS nesting between hole and electron pockets, their size-mismatch due to electron- or hole-doping reduce the effect of the FS nesting. Similar doping dependence of $T_{\rm SDW}$ in low doping levels for K and Mn suggests that the the size-mismatch of the FS is crucial for determining $T_{\rm SDW}$, irrespective to the type of dopant, until additional effects \emph{e.g.} structural modifications play a significant role at higher doping levels.

Concerning the superconducting phase in the electron-doping side, we found that both the magnetic and superconducting transitions are observed, suggesting the coexistence of two phases in the low doping regime. The doping dependence of $T_c$ does not exhibit a clear dome shape as found in Co-doped BaFe$_2$As$_2$,\cite{Ba122_Co:hhwen:hall} but shows a step-like increase of $T_c$ at low doping. In comparison with the polycrystalline data\cite{Sr122_Co:geibel:syn}, the Co-doping range for the superconducting phase is somewhat lower for single crystals. This discrepancy implies that the superconducting phase in doped-SrFe$_2$As$_2$ is sensitive to the synthesis method, which may generate \emph{e.g.} different internal stress of the samples. In fact, recently  Saha \emph{et al.} reported that even undoped SrFe$_2$As$_2$ can have a bulk superconductivity at 20 K\cite{Sr122:paglione:SC}. For the hole-doping side, K-doping induces the superconductivity above $x$ $\approx$ 0.2 while the superconducting phase is not observed for Mn-doping up to $x$ $\approx$ 0.4. Note that this doping level lies well inside the doping range where superconductivity is observed for K-doping.

Having established the phase diagram, we address why Mn-doping does not induce the superconductivity while Co- and K-doping do. For the electron-doping regime, it has been found that various doping with Co, Ni, Rh, and Pd results in exceptionally similar phase diagrams.\cite{Ba122_RhPd:Canfield:syn} They match very well with each other when plotted as a function of electron density added by those dopants. The number of extra electrons has been thus proposed as the control parameter of the overall phase diagram, in particular, for the superconducting phase in the electron-doping regime. This is, however, certainly not the case in the hole-doping regime as clearly seen by the distinctly different phase diagrams for K- and Mn-doping. In the hole-doping regime, the ground state is likely to be also determined by  several other parameters such as subtle structural modification of the FeAs layer and the rearrangement of charge density due to different nuclear charge of the dopants. Clearly, an explanation of the phase diagram in the hole-doped regime has to be sought beyond a rigid band picture.

\begin{figure}
\includegraphics*[width=8.5cm, bb=25 237 578 709]{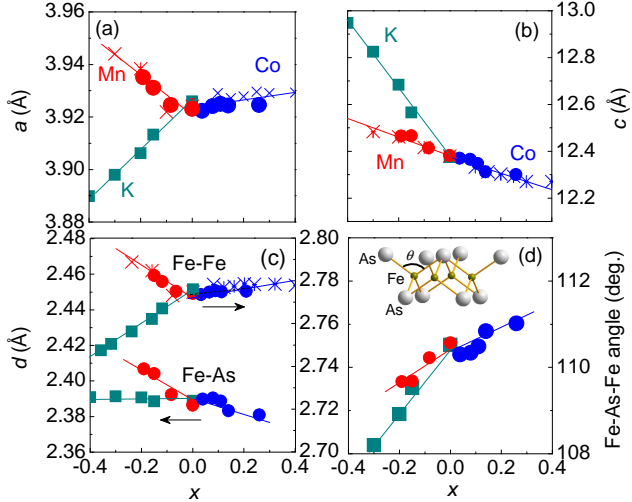}
\caption{\label{fig4}(Color online) Doping dependence of lattice parameters (a) $a$ and (b) $c$ for Co-, Mn-, and K-doping. The corresponding  bond lengths of Fe with neighboring Fe and As and  the bond angle around Fe with two neighboring As are shown (c) and (d), respectively. The local structure of the FeAs layer is shown in the inset of (d).  }
\end{figure}

Figure \ref{fig4} shows several structural parameters by different types of doping, K-, Co- and Mn-doping.\cite{note} Comparing the doping dependence of the lattice parameters for Mn- and K- doping, we note that the in-plane lattice parameter $a$ increases for Mn, but it is rapidly reduced for K. For the $c$-axis lattice parameter, both Mn and K doping result in its increase, but the increasing rate is much faster for K, which is partly understood as due to larger ionic radius of K$^{+}$ (1.37 $\rm{\AA}$) than Sr$^{2+}$ (1.18 $\rm{\AA}$). The corresponding Fe-Fe or Fe-As bond lengths are, therefore, quite different depending on the type of doping (see Fig. 4(c)). For Co-doping, the Fe-As distance decreases and the Fe-As-Fe angle becomes larger. Comparing Mn-doping with K-doping, the Fe-As-Fe angle is reduced in both cases, but with a somewhat larger rate for K-doping. In addition, K-doping reduces the Fe-Fe bond length with the almost constant Fe-As length, while Mn-doping increases the both Fe-Fe and Fe-As bond lengths.

The bond distances and bond angles are closely related to the effective hoping amplitudes between the neighboring sites. The increase of the Fe-As bond length with the reduced Fe-As-Fe bond angle would make the hoping path of Fe-As-Fe less effective for Mn doping than K doping. Recent electronic structure calculations\cite{FeAs:pickett:magnetism,FeAs:vildosola:As_z,FeAs:kuroki:As_z} suggest that the Fe-3$d$ bands with different orbital characters are entangled near the Fermi level, and its relative position and the band-width are very sensitive to the hopping amplitudes between nearest Fe neighbors. In particular, Fe 3$d_{x^2-y^2}$ bands, whose contribution to the density of states (DOS) is quite large due to its almost dispersion-less feature, is known to be easily shifted closer to the $E_F$ by reducing the nearest neighbor hopping. According to the Stoner criterion for itinerant magnets, formation of a magnetic moment is governed by the parameter $N(E_F)$$I$ where $I$ is a Stoner parameter, which is 0.7-0.9 eV for Fe. Therefore, increasing the Fe-As distance by Mn-doping is expected to favor the magnetic ground state because of the reduced Fe-As hopping and resultant larger contribution of narrow Fe 3$d_{x^2-y^2}$ bands. The calculated magnetic moment of SrFe$_2$As$_2$ is also expected to be increased as the Fe-As distance increases\cite{FeAs:pickett:magnetism}, which is in good agreement of the experimental observation. As shown in Fig. \ref{fig3}(b) the overall magnitude of the susceptibility increases with Mn doping in SrFe$_2$As$_2$.

The Mn$^{2+}$ itself, upon doping into SrFe$_2$As$_2$, is likely to favor a magnetic ground state. Mn$^{2+}$ has a half-filled $d$-shell with 3d$^{5}$ configuration, thus it has stronger Hund's coupling than Fe$^{2+}$ with 3$d^{6}$ configuration. The spin-polarized states, therefore, would be expected in the MnAs system. In fact, recent investigations on BaMn$_2$As$_2$ revealed an checkerboard-type AFM state with a relatively high $T_N$ $\approx$ 625 K and a semiconducting behavior.\cite{Ba122_Mn:neutron:goldman} Electronic structure calculations for BaMn$_2$As$_2$ also suggested strong spin-dependent hybridization between Mn $d$ and As $p$ states, leading to the AFM ground state.\cite{Ba122_Mn:band:singh}

\begin{figure}
\includegraphics*[width=8.5cm, bb=0 0 600 433]{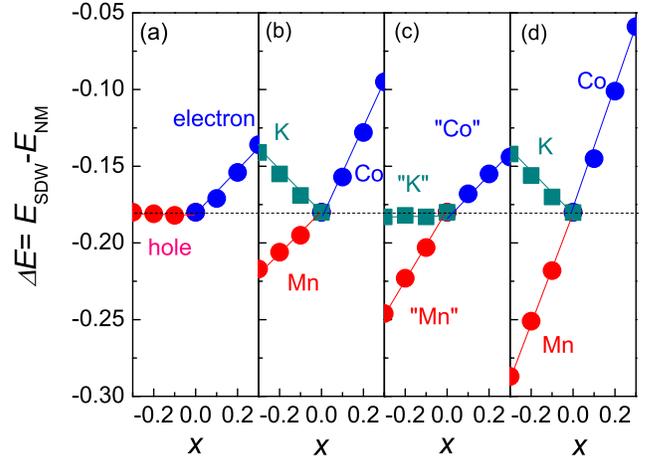}
\vspace{-3mm}
\caption{\label{fig5}(Color online) The calculated stabilization energies of the SDW states with respect to the nonmagnetic state for (a) SrFe$_2$As$_2$ with the shift of the chemical potential only, (b) K-, Mn- and Co-doped SrFe$_2$As$_2$ keeping the crystal structure same as pristine SrFe$_2$As$_2$, (c) undoped SrFe$_2$As$_2$ but assuming the crystal structure taken from K-, Mn- and Co-doped compounds, and (d) K-, Mn- and Co-doped SrFe$_2$As$_2$.}
\end{figure}

In order to elucidate a main deciding factor for superconductivity among several candidates such as the change of carrier counts, the structural modification and the chemical nature of the dopant, we calculate the total energies of the SDW phase using {\it ab-initio} calculations. In Fig. \ref{fig5} we present the stabilization energy, namely, the difference of the total energies ($\Delta E$) between those of the SDW phase ($E_{\rm SDW}$) and the nonmagnetic phase($E_{\rm NM}$). Thus, when -$\Delta E$ becomes smaller, the magnetic phase is suppressed, whereas the increase of -$\Delta E$ means that the magnetism gets stronger. For other magnetic structures such as ferromagnetic or checkerboard-type AFM phase show similar tendencies but with smaller stabilization energies (not shown).

First, we consider the doping effects only(Fig. \ref{fig5}(a)). In this case, we used the same crystal structure $i.e.$ the same lattice parameters and the same atomic position as SrFe$_2$As$_2$, except that $E_F$ is shifted corresponding to the doping level of Mn and Co. The stabilization energy of the SDW phase, -$\Delta E$ is significantly reduced for electron doping indicating suppression of the magnetic state. By contrast, for hole doping -$\Delta E$ remains almost constant. In terms of the Fermi surface nesting effects, both electron- and hole-doping spoil the nesting condition, thus relieving the electronic instability. However doping dependence of the electronic DOS ($N(E_F)$) is asymmetric between below and above $E_F$; the $N(E_F)$of SrFe$_2$As$_2$ strongly increases as the Fermi level ($E_F$) is lowered while $N(E_F)$ is rapidly reduced when the $E_F$ is moved to higher energies. This asymmetry originates from the fact that the electron pockets formed by the 3$d_{xz/zy}$ band with large bandwidth becomes prevails above $E_F$. On the other hand, below $E_F$, the electronic structure is dominated by a narrow Fe 3$d_{x^2-y^2}$ band region. Therefore in addition to the degradation of the nesting condition, the decrease of $N(E_F)$ in the electron doped system favors nonmagnetic states. For hole doping, however, the strong increase of $N(E_F)$ compensates the effects of degradation of the FS nesting conditions.

In the second step we consider the different dopant nuclear charge (see Fig. \ref{fig5} (b)). Here we used the same crystal structure as the former case in Fig. \ref{fig5}(a), but taking into account of the effects of different dopant using a virtual crystal approximation (VCA). Thus in this case, we consider the effects of the shift of the chemical potential plus the nature of dopant nuclear charge. We also performed the supercell calculations for $x$ = 0.25, which gave consistent results (not shown). For electron doping, -$\Delta E$ is further reduced as compared to considering simple doping effect in Fig. \ref{fig5}(a). For the hole doping side, different dopant gives different behavior; K-doping reduces -$\Delta E$, suppressing the SDW phase, while Mn-doping enhances -$\Delta E$, stabilizing the SDW phase.

\begin{figure}
\includegraphics*[width=8.5cm, bb=0 0 600 390]{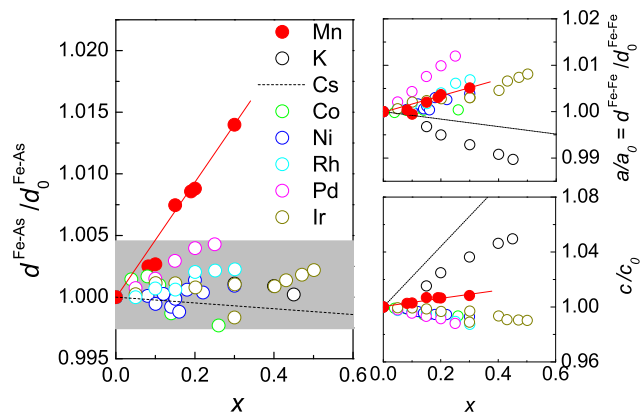}
\vspace{-3mm}
\caption{\label{fig6}(Color online) The doping dependence of (a) the Fe-As distance, (b) the Fe-Fe distance and (c) $c$-axis lattice parameter for Sr$_{1-x}$$A_x$Fe$_2$As$_2$ ($A$ = K and Cs) and SrFe$_{2-x}$$M_x$As$_2$ ($M$ = Mn, Co, Ni, Rh, Pd, and Ir) as compared to those of pristine SrFe$_2$As$_2$. Note that except Mn  (filled symbols), all other dopants (open symbols) induce superconductivity. }
\end{figure}

In order to compare with the effects of structural distortion only, we also calculate $\Delta E$ for hypothetical SrFe$_2$As$_2$ with its crystal structure taken from those of K-, Mn-, and Co-doped samples. In this case, we keep the $E_F$ same as the undoped compounds without any dopants. As shown in Fig. \ref{fig5}(c) -$\Delta E$ is significantly enhanced with the structure of Mn-doped samples while it stays almost constant for that of K-doped samples. This suggests that electronic structure, thus the ground state can be sensitively affected by even a slight change of the structure ($\sim$ a few \%). Combining all the effects of the chemical potential shift, the nature of dopant and structural distortion, $\Delta E$ as a function of doping is summarized in Fig. \ref{fig5}(d), which is roughly the sum of each contributions. For electron doping, most significant contribution for reducing the stabilization energy for the SDW phase comes from the shift of the $E_F$. The reduction of the Fe-As distance also makes a somewhat less, but comparable contribution. For the hole doping regime, the shift of the $E_F$ doesn't affect much, while the structural distortion, in particular, the increase of the Fe-As distance gives dominant contribution for stabilizing the SDW phase. We note that this behavior resembles the doping dependence of the Fe-As distance shown Fig. \ref{fig4}(c). This suggests that the Fe-As distance is the key parameter for the formation of the magnetic phase rather than other structural parameters, particularly for the hole doping regime.

Bearing this in mind, we checked the change of the Fe-As distance\cite{note} with various types of doping, K, Cs, Mn, Co, Ni, Rh, Pd and Ir.\cite{Sr122_K:sasmal:syn,Sr122_Ni:paglione:syn,Sr122_RhIrPd:hhwen:syn,SrNi2As2:bauer:syn,SrPd2As2:hofmann:syn,SrRh2As2:hellmann:syn} (see Fig. \ref{fig6}). These cover both electron and hole doping, different doping sites (A-sites and Fe-sites), different valence (1+, 1- and 2-), different period (3$d$, 4$d$, and 5$d$) as well as a wide range of ionic radius from 0.55 $\AA$ for Ni to 0.68 $\AA$ for Ir. The open symbols represent the dopants inducing superconductivity, while the solid symbols correspond to the dopants that results in the non-superconducting phase. For lattice parameters $a$ or $c$, their relative changes with respect to the pristine compound are quite random so that we cannot find any correlation with the (non)existence of superconductivity. In sharp contrast, for the Fe-As distance $d_{\rm Fe-As}$, it is clear that only Mn-doping that does not induce superconductivity leads to a strong increase of $d_{\rm Fe-As}$ while all the other types of dopant inducing superconductivity results in almost negligible (less than 0.5\%) variation of it. These findings clearly support our main conclusion that the increase of the Fe-As distance is detrimental for inducing superconductivity and favors the magnetic ground states.

In summary, using single crystals of a series of Co- and Mn-doped SrFe$_{2-x}$As$_2$, we explored the phase diagram in the hole and electron doping regimes. Both types of doping suppress the SDW transition in the lower doping regime, but at higher doping levels, the resulting ground state is quite different: nonmagnetic and metallic state with superconductivity for Co, but more magnetic and semiconducting state for Mn. The distinct physical properties of Mn-doped samples as compared to those of K-doped samples clearly demonstrate that for hole doping, the detailed nature of the dopant is important for inducing superconductivity. Based on the the first principle calculations, we show that slight doping-dependent structural change, in particular, the change of the Fe-As distance plays a key role for suppressing/stabilizing magnetic state. Such an exceptional sensitivity of the ground state to small structural changes suggests that modifying the electronic structure by structural distortions is more important than charge doping for inducing superconductivity.

\acknowledgments The authors acknowledge useful discussion with Bing Lv, and we thank E. Br\"{u}cher for experimental assistance. The work at SNU was supported by Korean government through NRL (M10600000238), GPP (K20702020014-07E0200-01410), basic science research programs (2009-0083512). The work at POSTECH was supported by the POSTECH Basic Science Research Institute Grant and by WCU through KOSEF (Grant No. R32-2008-000-10180-0). SHK was supported by Seoul R$\&$BD (10543).


\begin{thebibliography}{}
\bibitem{note} The Fe-As distance is determined by lattice parameters ($a$, $c$) and the atomic position of As ($z$). The $a$, $c$ and $z$ values for our samples were determined from the Reitveld refinements (Fig. \ref{fig1}). For other compounds whose lattice parameters are available in the literature, but not the $z$ value, we estimated $z$ using the linear interpolation between two end points, those of SrFe$_2$As$_2$ and the compound with full substitution for a considering dopant. Such a linear dependence of $z$ has been reported in various Fe pnictides, for example, (Ba,K)Fe$_2$As$_2$ (Ref. \cite{Ba122_K:johrendt:syn}).
\end{thebibliography}
\end{document}